# From Open Standards to Openly Governed

## Standards-Setting Organizations as Stewards of Openness amid Platformization and Digital Sovereignty

**Dr Begoña G. Otero and Dr Stefaan G. Verhulst**

The GovLab

**Authors' Note**

***Disclaimer***. The views and opinions expressed in this article are those of the authors and do not necessarily reflect the official position of The GovLab, New York University, or any of the organizations with which the authors are affiliated or from which they have received support. Any errors or omissions are the authors' own.

***Preprint status.*** This is a preprint. It has not been peer reviewed, and subsequent versions may differ in content. Please cite the version of record once available.

***Funding***. This research received no specific grant from any funding agency in the public, commercial, or not-for-profit sectors.

***Conflicts of interest***. The GovLab has undertaken advisory work for the Open Geospatial Consortium on its internal governance architecture. OGC is one of four organizations examined in this paper. All factual claims about OGC are supported by publicly available sources. The advisory engagement informed the authors' choice of governance questions and the practical calibration of the framework, but no non-public OGC material is cited or used to establish claims about OGC's governance performance. Otherwise, the authors declare no conflicts of interest.

***Use of generative AI***. The authors used Claude for literature screening and formatting (info viz and table formatting). The authors reviewed and take full responsibility for all content.

Correspondence concerning this article should be addressed to Begoña G. Otero at bgotero@thegovlab.org and Stefaan Verhulst at stefaan@thegovlab.org, The GovLab.

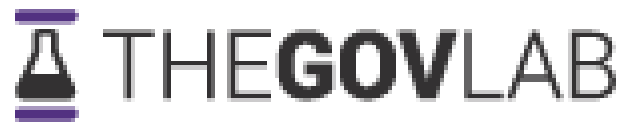

## Abstract

Open geospatial standards are publicly available, interoperable specifications that let geospatial data, services, and systems be shared, combined, and reused across platforms and organizations. That is the conventional definition, and it is where the difficulty begins: it describes a property of documents, not of the institutions that produce them or of the infrastructures in which they are implemented. Open standards are candidate digital public goods; once embedded in public systems they become components of digital public infrastructure. The first is a property of the artifact and its license, the second a role the artifact comes into play, and the two can come apart.

This paper asks two questions. Under what conditions do open geospatial standards function as digital public goods, and as digital public infrastructure, rather than as channels of enclosure? And which institutions are positioned to secure those conditions? We advance one claim in answer to each, and they are claims about different kinds of things. The first concerns institutional governance design, over which standards bodies have real control: the internal governance architecture of standards-setting organizations materially conditions whether their outputs hold their public-good character and continue to serve as infrastructure. The second concerns a development none of them chose. The geospatial field is being rapidly platformized, as proprietary location stacks, API governance models, and cloud-native Earth observation platforms shift effective standard-setting toward platform roadmaps, pricing structures, and service terms; sovereign-cloud and sovereign-AI initiatives relocate that authority rather than reversing it.

The two are joined by a feedback loop. Weak internal governance accelerates the migration of coordination authority to the implementation layer, and that migration erodes the incentive to invest in open standardization, creating the conditions under which platformization can capture the coordination functions standards bodies are meant to perform. The consequences reach beyond the geospatial domain, weakening practical portability and the conditions under which FAIR data can be combined and reused across systems. We do not argue that voluntary standards bodies are best placed to resist this; their leverage over the implementation layer is declining. The claim is narrower: they retain standing over the specification layer while remaining largely exempt from the participation and transparency obligations that bind formally recognized standardization bodies, and where law does not supply the constraint, internal governance must. We operationalize this through four recurring dimensions of governance, Purpose, Principles, People and Processes, and Policies and Practices (the 4Ps framework), offered as a means of sustaining the input, throughput, and output legitimacy on which the authority of these organizations depends, and we state its limit: these are instruments internal to a standards body, directed at a problem that is not.

***Keywords***: *geospatial standards, digital public infrastructure, digital public goods, standards-setting organizations, meso-institutions, governance, platformization, sovereign cloud, sovereign AI, Open Geospatial Consortium, W3C, IETF, Gaia-X, FAIR principles, digital twins, industrial metaverse, interoperability, open standards*, 4Ps framework, legitimacy.

## Table of Contents

# 1. Introduction: Open Geospatial Standards and the Governance Question

## 1.1. Open Geospatial Standards and the Meaning of Openness

Every time a weather radar overlay appears in a mobile app, or a traffic layer refreshes on a navigation screen, it reveals the quiet work of open geospatial standards: common ways to describe location, encode features such as roads and flood zones, and serve those layers over the web. Open geospatial standards are publicly available, interoperable specifications that enable geospatial data, services, and systems to be shared, accessed, combined, and reused across platforms and organizations. That definition spans two layers worth separating, because they are governed differently and on different timescales. For analytical purposes, this paper distinguishes a *canonical specification adopted through an SSO process*—the stable reference for what a compliant service or dataset should do—from the wider *implementation layer* of profiles, schemas, API descriptions, conformance artefacts, and implementation guidance against which developers actually build and which may evolve on a different timescale from the canonical specification.[1] The seamless layering of geospatial data that users take for granted depends on both. Openness here is a compound property, not a single one. The canonical specification must be publicly available. The wider implementation layer must remain implementable without permission or royalty where openness is claimed. And the process that produces and reviews both must be one that those affected can actually enter and contest. A standard can satisfy the first while failing the second and third; much of this paper is about what happens when it does.

The use of Geospatial information has become a cornerstone of digital public infrastructure,[2] underpinning climate adaptation, disaster risk management, urban planning, and a growing array of AI-enabled decision systems (UN-GGIM 2022; UN-GGIM Europe 2025; Kaleagasi 2022). Policy frameworks such as the United Nations Integrated Geospatial Information Framework (IGIF) and the UN-GGIM/OGC Standards Guide position open geospatial standards as enablers of FAIR data,[3] interoperable spatial data infrastructures, and Sustainable Development Goal-relevant services (UN-GGIM 2022; OGC 2024). The OECD's work on digital public infrastructure likewise emphasizes that DPI components such as identity, payments, data exchange, and geospatial infrastructures, should be built on open standards and digital public goods governed in the public interest (OECD 2024). Governance is not, on this conception, an external requirement imposed on such infrastructure but a part of it. As the G20 framework confirmed, governance is one of the three constitutive components of a DPI system, alongside technology and community. Our contribution works out what that component requires in the context of open geospatial standards.

[1] By “standards stack,” we mean not only canonical standards—the authoritative rulebooks agreed by a standards body—but also the specifications, schemas, profiles, APIs, conformance artefacts, and implementation guidance through which those standards are operationalized. Taken together, this stack determines how location-based data are classified, exchanged, validated, updated, and reused across institutions and sectors.

[2] Digital public infrastructure (DPI) is described in the G20 Digital Economy Ministers' consensus text as "a set of shared digital systems that should be secure and interoperable, and can be built on open standards and specifications to deliver and provide equitable access to public and/or private services at societal scale and are governed by applicable legal frameworks and enabling rules to drive development, inclusion, innovation, trust, and competition and respect human rights and fundamental freedoms": G20 Digital Economy Ministers' Meeting, *Outcome Document and Chair's Summary*, Bengaluru, 19 August 2023, para. 6.

## 1.2 The Governance Question

In much of the standards and policy discourse, however, the governance of open geospatial standards has been framed primarily in technical and procedural terms. Standards-setting organizations (SSOs)[4] have long provided the institutional backbone for technical standardization, and they are expected to ensure transparency, openness of participation, consensus-based decision-making, and permissive intellectual-property rules, thereby satisfying due-process principles articulated in trade, competition, and procurement law (Baron et al. 2019). Under this procedural view, the existence of open, consensus-based standards is treated as a sufficient condition for interoperability and for limiting vendor lock-in (Fiedler, Larraín, and Prüfer 2023). This paper argues that such a framing is becoming untenable.

The question is not whether open geospatial standards exist. It has two parts. First, under what conditions do open geospatial standards function as genuine digital public goods,[5]

---

[3] The FAIR principles, findability, accessibility, interoperability, and reusability, were formulated as guiding principles for scientific data management and stewardship. See Mark D. Wilkinson et al., "The FAIR Guiding Principles for Scientific Data Management and Stewardship," Scientific Data 3 (2016) 160018. FAIR is not a synonym for open. The principles are concerned with machine-actionability rather than with unrestricted availability, and they expressly contemplate restricted access, including authentication and authorisation where necessary, provided that the conditions of access are clearly specified and machine-readable. The distinction matters for the argument developed in Section 3.3: a platform may satisfy the formal FAIR criteria while effective openness comes apart.

[4] We use "standards-setting organization" (SSO) throughout as the umbrella term for bodies whose decisions set technical coordination rules, and avoid the common doublet with "standards development organization" (SDO), which marks a difference of literature convention rather than of substance. The distinction that does analytical work here is a different one: the degree of public-law recognition an organization's outputs enjoy. At one end sit the recognized standardization bodies whose outputs can become European or harmonized standards and, as such, acquire legal effects. Next are the consortium bodies examined in this paper (OGC, W3C, IETF) whose specifications gain legal traction only where they are separately referenced, identified, or incorporated by public authorities. Below them sit trust-framework operators such as Gaia-X, which set the conditions of exchange without producing standards at all, and finally the platform operators whose service terms and interface design constitute private ordering. Our argument concerns the migration of effective coordination authority down this gradient, and the term "standard-setting" is used on purpose: platforms set standards without developing any.

[5] The UN Secretary-General's Roadmap for Digital Cooperation does not itself define digital public goods; it calls on "all actors, including Member States, the United Nations system, the private sector and other stakeholders" to "promote open-source software, open data, open artificial intelligence models, open standards and open content that adhere to privacy and other applicable international and domestic laws, standards and best practices and do no harm." The operative definition adopted here is the Digital Public

open standards that are openly licensed, freely accessible, interoperable, and safe to reuse, and, once embedded in public systems, as digital public infrastructure, rather than as channels of enclosure? Second, which institutions are better positioned to secure those conditions, in an era of platform enclosure and digital sovereign infrastructure politics?

The stakes of this question extend well beyond technical coordination. Standards no longer simply coordinate engineers: they increasingly determine who can innovate, which markets remain contestable, what values become embedded in AI systems, whose knowledge is represented, and, ultimately, who exercises power over digital infrastructure. Because the outputs of standardization function as public goods and, once embedded in public systems, as public infrastructure, legitimacy can no longer derive from procedural openness alone, from the mere fact that "anyone can participate." Throughout this paper, accordingly, broader participation is treated not as an end in itself but as a mechanism for achieving legitimacy.

## 1.3 Two Claims

We advance two claims in answer, one to each part of the question, and they are claims about different kinds of things: one relates to institutional governance design, over which standards bodies have control; the other is a development in the field none of them chose.

**First**, voluntary SSOs in the geospatial domain should be understood as meso-institutional stewards of digital public goods: the open standards they produce are the candidate digital public goods, and, once embedded in public systems, function as digital public infrastructure; the SSOs themselves are the institutions whose internal rules and processes, and whose interactions with platforms and public authorities, materially condition whether those standards function as digital public goods or as channels of enclosure.[6]

**Second**, the increasing platformization of geospatial standard-setting, intensified by sovereign-cloud and sovereign-AI architectures, can compromise the practical public-good character and infrastructural relevance of open geospatial data standards, specifications and schemas. We do not claim that voluntary SSOs are the actors best placed to resist this: their leverage over the implementation layer is declining rather than growing. Our claim is narrower. These organizations retain standing over the specification layer, and they remain largely exempt from the participation and transparency obligations that bind formally recognized standardization bodies. Where law does not supply the constraint, internal governance must carry more of the burden, within its limits, and that requires their internal governance to be reviewed and updated so that they can future-proof themselves as legitimate stewards of the open geospatial standards stack.

---

Goods Alliance's, which builds on that language and adds relevance to the attainment of the Sustainable Development Goals (Digital Public Goods Alliance, Digital Public Goods Standard, 2025). We follow the Alliance's formulation while noting that it is a civil-society standard rather than an intergovernmental instrument. (UN Secretary-General, Roadmap for Digital Cooperation, 2020, 23, available at https://www.un.org/en/content/digital-cooperation-roadmap/assets/pdf/Roadmap_for_Digital_Cooperation_EN.pdf).

[6] On standards as 'invisible devices of sovereignty' that 'decide who can participate … and on what terms,' and as a hybrid public/private space exercising 'public power … through private … arrangements,' see Marta Cantero Gamito, 'Harmonising Consensus: The (Geo)Political Economy of Standardization in the AI Act,' 17 (2026) JIPITEC 75, 80, 84.

Both claims bear on legitimacy. The first concerns a condition of legitimacy, since governance is what produces or fails to; the second concerns a threat to the authority that legitimacy underwrites. Legitimacy is not, however, the outcome this paper is ultimately explaining. What is at stake is whether a standard holds its public-good character and, with it, its capacity to function as public infrastructure; and the two can come apart, since a specification can remain openly licensed while ceasing to be what anyone implements. Legitimacy sits between governance and that outcome: it is what allows a standards body to hold coordination authority at all. Thus the asymmetry that runs through the paper. A platform holds coordination authority through market position, installed base, and switching costs, and needs no one's consent to exercise it; a standards body holds it only by the continuing consent of those its standards bind.

The two claims are connected by a hypothesized feedback mechanism. Weak internal governance degrades the fit between what a standards body publishes and what implementers build against; implementers may then default to whatever interfaces are available and maintained, which increasingly means platform interfaces; and as coordination authority accretes at the implementation layer, the incentive to invest scarce engineering time in open standardization may fall further. Recent foresight work on data governance describes a related dynamic: where openness is experienced as exposure to extraction rather than reciprocal contribution, contributors may restrict what they release (Zable and Verhulst 2026). Platformization is therefore not only an external pressure on SSOs but partly a product of how they are governed. Stated as a proposition: weak internal governance creates the institutional conditions under which platformization can more easily capture the coordination functions that SSOs are meant to perform. The cases in Section 2.2 and the platform histories in Section 3 make this mechanism plausible and specify its steps; yet, they do not establish the causal sequence.

## 1.4 Contribution, Scope and Limitations

Our contribution draws on earlier governance-infrastructure research and on The GovLab's advisory engagement with the Open Geospatial Consortium on its internal governance architecture. That engagement forms part of the epistemic basis of this paper rather than merely its background: it is where several of the recurring patterns described in Section 4 were first observed. Two constraints follow, and we state them rather than obscure them. First, the engagement was advisory rather than a research study. It yielded practice-derived knowledge, not systematic data, and the generalizations in Section 4 should be read as informed judgment rather than as findings a reader can independently verify. Second, its confidential material is not reproduced here: every statement of fact about OGC is sourced to public documents, and observations from the engagement appear only as unattributed patterns, never as findings about any single organization.

Three limits bound what follows, and are best stated at the outset. The first concerns evidence and method. This is a normative-conceptual and documentary analysis, not an empirical comparative case study. The four organizational cases are analytically selected as contrastive illustrations of distinct governance problems and institutional responses, and are read through two lenses specified in advance: input, throughput, and output legitimacy, and the 4Ps governance framework. They are not a sampled population, we do not claim representativeness, and no causal inference is drawn from them. Nor do we test whether the reforms examined have achieved the outcomes for which they were designed..

The second concerns scope. Our cases are consortium bodies and one trust-framework operator, all North American or European. We do not examine the formally recognized standardization organizations whose outputs carry legal effects, although Section 5 argues that the legitimacy demands now being made of them are migrating toward consortia that lack the corresponding obligations; nor do we examine standardization in China, India, or the regional bodies through which much of the world's geospatial capacity is being built, despite the claims we make about global legitimacy and about who bears the consequences of standards.

The third concerns the reach of the remedy, and is the most significant. The reforms proposed here are internal to standards bodies, and deliberately so: they address the part of the problem those bodies can act on directly. Other conditions also shape whether an open standard remains open in practice: The legal conditions under which a compatible implementation may lawfully be built, the competition-law treatment of dominant interfaces, or the procurement and recognition decisions of public authorities, and all lie beyond the reach of any standards body acting alone. We return to this limit in Section 4. It means that the framework offered here is necessary rather than sufficient, and that a full account of governable openness would require a companion analysis of the legal and market conditions that this paper does not attempt, and further research in this field is needed.

The paper proceeds as follows. Section 2 situates SSOs as meso-institutions and introduces the two analytical lenses used throughout. First, the three dimensions of legitimacy, and second, our four dimensions of governance (Purpose, Principles, People and Processes, and Policies and Practices, the "4Ps model"[7]) before assessing four organizational cases, the OGC, W3C, IETF, and Gaia-X, through both. Section 3 analyzes how platformization and sovereign-AI stack dynamics reconfigure effective standard-setting, and sets out the risks these pose for FAIR data practices, digital twins, and the industrial metaverse. Section 4 develops the 4Ps into a diagnostic framework, proposes a Public Interest Test to accompany standards across their lifecycle, and states what a standards body cannot secure on its own. Section 5 concludes.

## 2. Standards-Setting Organizations as Mes-Institutions: Internal Governance Architectures

### 2.1 The Meso-Institutional Framework for Standards Governance

[7] The 4Ps model has been developed in the realm of data governance, offering a practical, rights-based guide and more importantly a self-assessment framework to help organizations evaluate their current capabilities. (see: Stefaan Verhulst, Data Governance Toolkit, 2025). This paper adapts that framework to standards-setting organizations and combines it with legitimacy criteria drawn from established good-governance scholarship.

Economic and management scholarship has long recognized that standards are not merely technical artifacts but politico-economic institutions[8]. Compatibility standards can reduce transaction costs, enable network externalities, and facilitate a division of innovative labor, yet they can also create lock-in and durable market power (Katz and Shapiro 1985; Shapiro and Varian 1999). In digital and platform markets, standards function as boundary resources that structure how third parties access data, users, and functions, and thereby shape the evolution of whole ecosystems.

Recent work in institutional economics offers a particularly productive lens for understanding the governance dimension of voluntary standards. This work conceptualizes voluntary standards as *meso-institutions*: coordination systems that occupy the analytical space between micro-level transactions and macro-level law (Soregaroli et al. 2022). In their framework, meso-institutions perform three key functions: (i) translating and adapting broad rules and norms to specific contexts and sectors; (ii) monitoring their implementation; and (iii) enforcing regulations. Their empirical investigation of private standards in the European soybean supply chain demonstrates that firms participating in a private standard exhibit fewer hybrid governance forms and higher levels of spot-market transactions, suggesting that the meso-institution fulfills some of the coordination functions otherwise performed by more costly bilateral arrangements. The implication for geospatial standardization is direct: the SSOs that develop and maintain open geospatial standards act as meso-institutions that translate general normative commitments (openness, interoperability, public interest) into specific mechanisms (membership rules, decision procedures, lifecycle processes, external collaborations, research and development among members and compliance frameworks) that delineate the domain of permissible and expected activities for heterogeneous actors across the geospatial ecosystem. Whether that translation yields a genuine digital public good, that is, an open standard that is openly licensed and freely accessible, interoperable, and safe to reuse, depends on a small set of governance choices that recur across organizations, which we develop as the 4Ps framework in Section 4.[9]

This meso-institutional perspective aligns with and extends broader scholarship on SSO governance. On the one hand, those who model voluntary SSOs as self-governed clubs of innovators and implementers and show that internal rules about the relative power of implementers shape participation incentives, royalty levels, standard quality, and welfare (Fiedler, Larraín, and Prüfer 2023). Their central result, an inverted-U relationship between implementer control and welfare, underscores that SSO statutes (membership categories, voting thresholds, intellectual property policies) are key parameters in the political economy of standards, not mere procedural details. On the other hand, others have empirically demonstrated that the affiliations of individuals in leadership positions—working group chairs in particular—influence the direction of standards development, even when formal procedures appear neutral (Baron and Kanevskaia 2023). Their analysis also reveals that while attendee diversity has increased in many global standards' organizations, traditional incumbent stakeholder categories and Western nationals continue to occupy a disproportionate share of managerial positions. These findings confirm that the internal governance architecture of an SSO is a material determinant of whether the standards it produces serve broad public interests or reflect the preferences of dominant participants.

[8] Standards are ‘a microcosm in which conflicting epistemologies, regulatory philosophies, national traditions, social values, and professional attitudes are faithfully reflected’: Annalisa Volpato, Mariolina Eliantonio and Michèle Finck, ‘Harmonised Standards in AI Regulation’ (Editorial), 17 (2026) JIPITEC 1, 5 (quoting Majone); see also Elvira M.R. Oliva, ‘Trustworthy AI Through Standards?,’ 17 (2026) JIPITEC 61, 64-67 (‘technicalities cannot be separated from normative values’).

[9] These criteria track the openness, open-licensing, interoperability, and do-no-harm indicators in Digital Public Goods Alliance, Digital Public Goods Standard (2025).

Legal scholarship on transnational private regulation supports the view that standard-setting bodies may exercise regulatory functions through instruments such as technical standards, certification schemes, and codes of conduct that are subsequently recognized, referenced, or embedded in public law, thereby generating hybrid configurations of public and private authority (Cafaggi 2011). This literature further suggests that governance outcomes are often shaped through the interaction of overlapping normative layers, including corporate rules, voluntary standards, and formal regulation, with the result that ostensibly voluntary standards may acquire quasi-mandatory force where they condition access to markets, supply chains, or public procurement (Bartley 2011; Cafaggi 2011). In the geospatial domain, this dynamic is visible in frameworks such as INSPIRE, which incorporates OGC and ISO standards as binding elements of European spatial data infrastructure, and in national SDI programs that treat particular geospatial specifications as foundational components of public digital systems (Agrawal 2025).

Related work on meso-institutions in sustainability transitions provides a useful analytical extension by showing how intermediate organizations mediate between macro-level governance frameworks and micro-level implementation, translating broad policy goals into operational arrangements (see e.g. de Mello et al. 2024). Applied to geospatial standardization, this suggests that SSOs should be understood not simply as technical coordinators, but as institutional intermediaries that render public-interest commitments such as interoperability, openness, and data as digital infrastructure, operational through specific procedural, technical, and organizational forms.

These findings are easier to make sense of through the literature on legitimacy, which asks a central question: what makes a rule-making body worthy of authority and trust? A common answer is that legitimacy rests on three dimensions (Scharpf 1999; Schmidt 2013; Werle and Iversen 2006):

- First, there is *input legitimacy*: who gets to participate, whose interests are represented, and who is missing from the table.
- Second, there is *throughput legitimacy*: how decisions are made, whether procedures are transparent and fair, whether conflicts of interest are disclosed, and whether accountability mechanisms work when processes fail.
- Third, there is *output legitimacy*: whether the standards produced actually solve coordination problems, deliver interoperability and public value, avoid capture by dominant interests, and earn the confidence of implementers and affected communities.

For voluntary SSOs, these three dimensions do not automatically reinforce one another. An organization may appear open in terms of participation while still relying on opaque internal procedures or producing outcomes that mainly reflect the preferences of its most powerful members. In our own work with standards bodies, we have therefore found it more useful to treat legitimacy not as a rhetorical commitment but as something that can be observed and tested in practice, through transparency, participation, accountability, agility, and technological responsiveness. This legitimacy lens runs through the rest of the paper: the case studies in Section 2.2 can be read as attempts to address specific legitimacy deficits, while Section 4 turns those concerns into a practical governance review framework.

Two analytical lenses therefore run through the rest of the paper. The first is the legitimacy trichotomy just set out. The second is a model of four recurring dimensions of governance, adapted from data-governance scholarship and calibrated here to voluntary standards bodies: Purpose asks what and whose public value a standards body exists to serve; Principles concern the commitments that should inform and constrain how it acts; People and Processes concern who participates, who decides, and how decisions can be challenged; and Policies and Practices concern how those commitments are translated into operational routines across the standards lifecycle. We refer to these as the 4Ps framework. The two lenses answer different questions and are complementary rather than alternative: legitimacy specifies what must be secured, while the 4Ps locate where within an institution it is produced or lost.

**Legitimacy and the 4Ps Framework in Technical Standardization**

**Input legitimacy**
*Who participates?*
- Who participates?
- Who is missing?
- Are public-interest actors represented?

**Throughput legitimacy**
*How are decisions made?*
- Are decisions transparent?
- Are conflicts disclosed?
- Are deliberations fair?
- Is there accountability?

**Output legitimacy**
*What results are produced?*
- Do standards actually serve interoperability?
- Do they increase public value?
- Do they avoid capture?
- Are they trusted?

**Legitimacy of rule-making institutions**
*What makes a standards body worthy of authority and trust?*

These legitimacy questions are operationalized through the 4Ps governance framework.

**The 4Ps framework**

**Purpose**
(WHY)
What public value the standard and its steward exist to create, and for whom
*chiefly output legitimacy*

**Principles**
(HOW)
Which commitments are non-negotiable: open licensing, free public access, interoperability, and do-no-harm
*chiefly throughput legitimacy*

**People and Processes**
(WHO)
Who participates, how decisions are made, and how they are contested
*chiefly input legitimacy*

**Policies and Practices**
(WHAT)
What the organization actually does over time across the standards lifecycle
*chiefly output legitimacy*

The 4Ps connect legitimacy to observable governance design and day-to-day operation.

*These questions structure a judgment about legitimacy; they do not substitute for one.*

Conceptual basis: Scharpf (1999); Schmidt (2013); Werle and Iversen (2006); Verhulst (2025); Gonzalez Otero and Verhulst (2025).

The meso-institutional framing carries a critical governance implication. If SSOs are meso-institutions that mediate between broad norms and specific practices, then their internal governance architectures (how they define membership, allocate decision rights, manage conflicts of interest, steward the lifecycle of standards, and ensure accountability) are not peripheral organizational details. They are among the mechanisms through which the public-good character of geospatial standards is either maintained or eroded[10]. Weak internal governance, normally characterized by concentrated participation, opaque decision-making, slow adaptation, or insufficient accountability, does not merely reduce organizational efficiency. It creates the institutional conditions under which platformization can more easily capture the coordination functions that SSOs are meant to perform. These architectures are, equally, the mechanisms through which legitimacy is produced: a meso-institution's authority ultimately rests not on formal openness alone, but on whether its translation of broad norms into specific rules is perceived as legitimate by those who bear the consequences of the resulting standards.

The above legitimacy framework identifies what must be secured—meaningful participation, fair and accountable processes, and public-value outcomes. In Section 4 we suggest 4Ps as a complementary institutional lens for identifying where those conditions are produced. People and Processes are central to input legitimacy; Principles and procedural safeguards contribute to throughput legitimacy; and Purpose together with Policies and Practices shape whether output legitimacy is ultimately achieved. The cases that follow illustrate how weaknesses in each of these dimensions generate different forms of governance deficit.

## 2.2 Meso-Institutional Adaptation in Practice: OGC, W3C, IETF, and Gaia-X

The meso-institutional framework is not only a theoretical lens; it offers a way to read ongoing governance reforms across several major SSOs as deliberate efforts to strengthen or rehabilitate the translation and coordination functions that justify these organizations' institutional roles. Four cases from the geospatial and adjacent standards landscape illustrate the pattern.

### *2.2.1 The Open Geospatial Consortium: Broadening Participation and Modernizing Outputs*

The Open Geospatial Consortium (OGC) is an international, non-profit standards organization founded in 1994 that brings together governments, industry, research institutions, and the geospatial community to advance the interoperability of location-based data and technologies. Since late 2025, OGC has undertaken a series of governance-related reforms that, read through the meso-institutional lens, represent deliberate efforts to strengthen the organization's capacity to translate broad openness commitments into operational governance practices.

[10] The point that opaque private rule-making transfers 'indirect normative power … removed from the institutional control mechanisms … proper to a state governed by the rule of law' is developed, for harmonised standards, by Mariolina Eliantonio, 'The Role of the European Commission in the Control of Harmonised Standards under the AI Act and the Challenges of "Opacity",' 17 (2026) JIPITEC 49, 54-55. On the absence of a 'division of power between those who produce … and those who … evaluate risks,' both left to industry-dominated standardisation bodies, see Simone Casiraghi and Niels van Dijk, 'The Politics of Standardising Ethics,' 17 (2026) JIPITEC 36, 38, 44.

*Individual membership and participation architecture.* In March 2026, at its 134th Member Meeting in Philadelphia, OGC officially launched an Individual Membership category.[11] This reform expanded the participation architecture beyond its traditional model of organizational membership, creating a direct pathway for independent developers, consultants, researchers, and students to participate in working groups, code sprints, testbeds, and ongoing collaboration. The pricing structure was designed on a sliding scale linked to the World Bank country income classification, adjusting automatically based on country of residence. This reform addresses a well-documented weakness of voluntary SSOs: the tendency for participation to concentrate among well-resourced corporate and governmental actors, thereby skewing the "translation" function of the meso-institution toward the preferences of dominant players.

*The Agora collaboration platform.* In 2025, OGC launched Agora, a member engagement platform designed as the primary online space for member collaboration.[12] Agora integrates working-group coordination, discussion forums, document sharing, and committee-activity tracking into a single environment accessible on mobile and desktop. In governance terms, Agora represents an attempt to lower the transaction costs of participation (a core function of meso-institutions) by reducing the informational and logistical barriers that have historically made SSO engagement difficult for actors outside the core institutional membership.

*API modernization and the Building Blocks framework.* OGC has made substantial progress transitioning from legacy Web Service Standards (WMS, WFS, WCS, WPS) to a new generation of OGC API Standards following modern RESTful web development practices. The publication of the OGC API – Connected Systems Standard in 2025, characterized as a "modernization and realignment" within OGC's Building Blocks framework, exemplifies this shift (OGC 2025). Code sprints and innovation testbeds have accompanied this technical modernization, including a hybrid code sprint in January 2026. Institutionally, this modernization is significant: if the SSO's outputs no longer align with how developers actually build, the coordination function migrates to whatever actor does offer implementable interfaces, which increasingly means platform operators.

*Governance as an explicit agenda item.* Governance improvement has been a recurring agenda item across OGC's recent convenings rather than a single-meeting concern. It was taken up at the 132nd Member Meeting in Mérida, Yucatán, Mexico (June 9–12, 2025), and again at the 133rd OGC Member Meeting in Boulder, Colorado (October 28–31, 2025), where it was explicitly identified as a key discussion topic. The meeting also focused on how OGC standards can support appropriate use of artificial intelligence, a theme continued at OGC iDays 2025 in Bad Nauheim, Germany (December 9–10, 2025).[13] That governance has moved from an implicit organizational concern to an explicit agenda item at OGC's highest-level convenings signals institutional recognition that the procedural and structural dimensions of standard-setting require deliberate attention. Broadening participation in this way strengthens the people-and-processes dimension developed in Section 4, and the inclusivity a digital public good requires. Read through the legitimacy lens, the sliding-scale membership and the Agora platform are, moreover, capacity-building measures rather than merely door-opening ones: they lower the financial and informational barriers that determine whether formally open participation becomes actual participation.

[11] OGC, "OGC Individual Membership Officially Launches," OGC Blog, March 25 2026, https://www.ogc.org/blog-article/ogc-individual-membership-launch/ . Individual Membership was officially launched at the 134th OGC Member Meeting in Philadelphia (March 2–5, 2026). Individual members receive OGC Agora licenses, complimentary registration for one in-person meeting per year, and may participate in working groups and code sprints. Pricing is based on a sliding scale linked to the World Bank country income classification. See also: OGC, "OGC Individual Membership: Join the Global Geospatial Community," 2026, https://www.ogc.org/membership/individual/.

[12] Agora is OGC's primary online space for member collaboration, where working groups coordinate, discussions take place, documents are shared, and members connect across projects and regions. Through Agora, members engage in working groups and innovation initiatives, collaborate with other members, and join compliance discussions. See: OGC Agora, Google Play, https://play.google.com/store/apps/details?id=org.ogc.agora.

#### *2.2.2 The World Wide Web Consortium: Reconstitution as a Public-Interest Non-Profit*

The World Wide Web Consortium (W3C) is an international standards organization that develops open technical standards and guidelines for the World Wide Web. It provides a particularly instructive case of meso-institutional restructuring. For its first quarter-century, W3C existed not as a single legal entity but as a set of contractual agreements among four host universities: MIT (United States), ERCIM (France), Keio University (Japan), and Beihang University (China). This arrangement, while operationally effective during the early web era, created governance limitations: decision rights were distributed across multiple institutional hosts, accountability structures were diffuse, and the organization's legal capacity to act as a unified governance actor was constrained.

On January 31, 2023, W3C was reconstituted as W3C Inc., a public-interest non-profit organization incorporated under United States law with 501(c)(3) status.[14] The reconstitution involved several governance design choices that are analytically significant. A Board of Directors was seated with a member-elected majority: of the initial eleven seats, seven were elected by the W3C Membership to bring a multi-stakeholder perspective, while four represented the founding Host institutions to ensure global geographic focus.[15] As a 501(c)(3) entity, W3C is now subject to mandatory transparency requirements: publicly available bylaws, annual financial audits by independent external auditors, and disclosure of tax documents and operating statements.

[13]The 133rd OGC Member Meeting (Boulder, Colorado, October 28–31, 2025) focused on emergency management and marine geospatial data, with key topics including extreme weather and use of artificial intelligence within OGC; governance improvement was also identified as a key topic of discussion. The OGC iDays 2025 (Bad Nauheim, Germany, December 9–10, 2025) included discussion of actions OGC should take to ensure its standards can support appropriate use of artificial intelligence. See: OGC, "OGC iDays 2025," https://events.ogc.org/InnovationDaysFrankfurt.

[14]World Wide Web Consortium, "W3C Re-Launched as a Public-Interest Non-Profit Organization," press release, January 31, 2023, https://www.w3.org/press-releases/2023/w3c-le-launched/. The new entity preserves the member-driven approach and existing worldwide outreach while allowing for additional partners around the world, and preserves the core process and mission of the Consortium to shepherd the web by developing open web standards as a single global organization.

The W3C case also carries a caution that the reconstitution narrative alone would obscure. On 6 July 2017 the Director decided to publish Encrypted Media Extensions as a Recommendation, over a sustained objection led by the Electronic Frontier Foundation that the specification would expose security researchers and interoperable implementers to liability under anti-circumvention law (section 1201 of the United States Digital Millennium Copyright Act and its European analogues) unless members granted a covenant not to sue. The EFF appealed on 12 July, in what it described as the first appeal against the Director in W3C's history; the appeal secured the membership support required to proceed to a vote of the Advisory Committee, which sustained the Director's decision by 108 votes to 57, with 20 abstentions. EME was published as a Recommendation on 18 September 2017, and the EFF resigned from the Consortium the same day (W3C 2017; Electronic Frontier Foundation 2017). The episode is instructive precisely because nothing procedural went wrong: participation was open, the process was followed as written, an appeal mechanism existed and was used, and the membership voted. What the process could not accommodate was a substantive objection about who would bear the risks of implementation, raised by a participant with no market power to trade. Read against Section 4, this is the failure a Public Interest Test is designed to surface: EME satisfied the three questions standards bodies routinely ask: market demand, technical consensus, implementation experience. It failed the two they do not, namely who bears the risks and which constituencies were absent. The reconstitution described above should not be read as the remedy for this deficit. It addressed a different one, and structural reform of legal personality and fiduciary accountability does not by itself resolve throughput legitimacy.

Read through the meso-institutional lens, W3C's reconstitution represents a deliberate strengthening of the governance capacities, from legal personality, fiduciary accountability, transparent decision-making, to member-driven strategic authority, that underpin the organization's ability to function as a credible coordination institution. The transition preserved the core consensus-based process and mission while creating the institutional infrastructure needed to sustain that mission in a more contested standards landscape. For the geospatial community, the W3C case is relevant because it demonstrates that meso-institutional adaptation can involve fundamental structural transformation, not merely incremental procedural reform. It also illustrates that when an SSO's governance architecture is perceived as structurally inadequate, the response may need to be constitutive, that is, redesigning the organization's legal and governance foundations, rather than operational. In the terms of 4Ps framework (Section 4), this reinforces the principles dimension, open, accountable governance, on which a standard's public-good status ultimately rests.

[15]The W3C Board of Directors—first seated in September 2022—is the governing body of the World Wide Web Consortium public-interest non-profit organization, with ultimate authority on W3C's strategic direction, legal obligation to ensure W3C implements its mission, and fiduciary responsibility over W3C as a whole. The initial composition comprised 11 members: 7 voting seats elected by the W3C Membership to bring a diverse multi-stakeholder perspective, and 4 voting seats representing the Host institutions (MIT, ERCIM, Keio, Beihang) to ensure a global focus. See: W3C, "First Board of Directors to Initiate Critical Functions of W3C Inc.," W3C News, September 2022, https://www.w3.org/news/2022/first-board-of-directors-to-initiate-critical-functions-of-w3c-inc/. See also: W3C, "Corporation," About Us, https://www.w3.org/about/corporation/.

### *2.2.3 The Internet Engineering Task Force: Diversifying Participation in a Rough-Consensus Model*

The Internet Engineering Task Force (IETF) is an international standards organization that develops open technical standards for the architecture and operation of the Internet. It presents a different but complementary case. As the principal SSO for Internet protocols, the IETF operates through a distinctive "rough consensus and running code" model that emphasizes technical merit and open participation over formal voting or institutional membership. However, research on IETF participation patterns reveals persistent challenges. Studies examining two decades of IETF affiliations demonstrate that while attendee diversity initially grew, it has subsequently remained broadly constant or even declined in some dimensions (Zhang et al. 2025). Organizational diversity and geographic representation remain significant concerns.

The IETF's governance response has involved several reforms relevant to the meso-institutional framework. RFC 9389, published in 2023, updated the longstanding RFC 8713 rules governing the Nominating Committee (NomCom), the body responsible for selecting IETF leadership, including Area Directors, the IAB, and the IETF LLC.[16] The reform expanded eligibility criteria to include remote meeting attendance, explicitly acknowledging that the previous requirement for in-person attendance at three of the past five meetings created barriers to participation from underrepresented regions and organizations. Additionally, an Internet-Draft titled "The IETF Is for Everyone: Toward Inclusive and Equitable Participation in Internet Governance" proposes collaborative strategies including mentorship, multilingual onboarding, university engagement, and periodic community consultations to align IETF practices with its foundational commitments to openness and global reach.[17]

For the geospatial standards community, the IETF case illustrates two important dynamics. First, even SSOs with strong procedural commitments to openness can develop participation patterns that systematically underrepresent significant constituencies, a form of meso-institutional drift in which the translation function becomes biased over time. Second, governance reform in the IETF involves both formal rule changes (RFC updates) and informal cultural interventions (mentorship, onboarding, community consultations), confirming that meso-institutional adaptation operates simultaneously at procedural and practice levels. The deficit it addresses is one of people and processes: a standard is not a genuine public good if those it affects cannot meaningfully shape it. Like OGC's reforms, the IETF's remote-eligibility rules and mentorship initiatives are investments in the capacity to participate, not merely in the formal right to do so.

---

[16]RFC 9389 updates RFC 8713 by defining a new set of eligibility criteria for the IETF Nominating Committee from first principles, with consideration to the increased salience of remote attendance. Under the updated rules, an individual may serve on the NomCom if they have attended at least 3 out of the last 5 IETF meetings (in person or online), or have been a Working Group Chair or Secretary within the last three years, or have been listed as an author or editor on at least two IETF Stream RFCs within the last five years. See: IETF, "Nominating Committee Eligibility," RFC 9389, April 2023, https://www.rfc-editor.org/rfc/rfc9389.html.

[17]Attoumani, M., et al., "The IETF Is for Everyone: Toward Inclusive and Equitable Participation in Internet Governance," Internet-Draft, draft-attoumani-ietf-inclusion-03, 2025, https://www.ietf.org/archive/id/draft-attoumani-ietf-inclusion-03.html. Through collaborative strategies such as mentorship, multilingual onboarding, university engagement, and periodic community consultations (e.g., Africa IGF 2025), this draft invites discussion on how the IETF can better align its practices with its foundational commitments to openness and global reach.

#### *2.2.4 Gaia-X: From Manifesto to Operational Trust Framework—and Its Geospatial Implications*

Even though Gaia-X is not an SSO per se, it is included here precisely for that. Formally, Gaia-X European Association for Data and Cloud (AISBL) is a non-profit association that develops federated frameworks and specifications for trusted and interoperable data and cloud ecosystems and operates a compliance and trust-labeling scheme. It does not produce standards through an accredited or consortium standardization process; where its design principles have entered formal standardization, they have done so through CEN/CENELEC rather than through Gaia-X itself. Yet, it offers perhaps the most directly consequential case for geospatial standardization, because it operates at the intersection of standards governance, data-space architecture, and sovereign-infrastructure policy, which is precisely the terrain where the governance-infrastructure gap seems most acute.

Launched in 2019 as a Franco-German initiative for European digital sovereignty, Gaia-X has evolved from a conceptual manifesto into an operational framework governing data exchange, trust, and interoperability across European data spaces. The Danube release of the Gaia-X Trust Framework (version 3.0), published in November 2025, introduces domain and geographic extensions that enable trust to be federated across diverse ecosystems and compliance regimes.[18] Since spring 2025, the Gaia-X design principles have been incorporated into CEN/CENELEC Trusted Data Transaction (TDT) standards, positioning them as part of the formal European standardization architecture.[19] More than fifteen European data spaces have now emerged with developing standards, legislative support, and federated tooling (Gaia-X 2025).

The geospatial implications are significant and underexplored. Gaia-X's Data Exchange Services specification defines vocabulary, conceptual and operational models, policies, and ontologies for data exchange designed to deliver trust, interoperability, discoverability, and traceability. These are precisely the governance functions that geospatial data standards are also designed to perform, but at a different layer of the stack. When geospatial data and services are deployed within Gaia-X-compliant data spaces, the governance rules applicable to those data and services are shaped not only by OGC or ISO standards but also by Gaia-X trust labels, compliance automation mechanisms, and the specific sovereign-cloud infrastructure within which the data space operates.

[18]Gaia-X European Association for Data and Cloud, "Danube Release: Trust Framework 3.0," November 2025, https://gaia-x.eu/gaia-x-enters-season-two-of-dataspaces-and-digital-ecosystems-with-summit-2025/. The Danube release introduces domain and geographic extensions, enabling trust to be federated across diverse ecosystems and compliance regimes. Higher trust levels are being positioned as compatible with forthcoming EUCS "High+" requirements. More than 15 European data spaces have emerged with developing standards, legislative support, and federated tooling.

[19]Since spring 2025, the Gaia-X design principles have been incorporated into CEN/CENELEC Trusted Data Transaction (TDT) standards, positioning them as part of the formal European standardization architecture. Gaia-X also published the Data Exchange Services specification, which defines the vocabulary for data exchange, sets the definition of Data Exchange Services as well as conceptual and operational models, data exchange policies, and ontologies for data exchange to deliver trust, interoperability, discoverability, and traceability to the data economy. See: International Data Spaces Association, "Gaia-X's Data Exchange Services Specifications: A Milestone for Sovereign Data Sharing in Europe," 2025, https://internationaldataspaces.org/gaia-xs-data-exchange-services-specifications-a-milestone-for-sovereign-data-sharing-in-europe/.

Recent developments confirm this intersection. Esri has confirmed that ArcGIS architecture aligns with Gaia-X requirements, enabling decentralized collaboration that protects data sovereignty (Esri 2025).[20] Airbus aims to establish an aerospace data space by 2026 using Gaia-X frameworks (Datacenter Dynamics 2024). X-Road 8 (“Spaceship”), scheduled for 2026, will make the X-Road Trust Framework interoperable with Gaia-X specifications (NIIS 2025)—a development with direct implications for cross-border spatial data infrastructures in the Nordic and Baltic regions.

From a meso-institutional perspective, Gaia-X represents a new type of governance intermediary: not a traditional SSO developing technical specifications through consensus, but a trust-framework operator that sets the conditions under which data exchange, including geospatial data exchange, can occur within sovereign-compliant environments. The analytical question for the geospatial community is whether this new meso-institutional layer complements or competes with existing SSO governance. If Gaia-X trust labels, compliance frameworks, and data-space architectures become the operative governance rules for geospatial data in Europe, therefore determining who can access what data, under what conditions, with what provenance and quality guarantees, then the effective standard-setting function has partially migrated from geospatial SSOs to Gaia-X’s governance infrastructure, regardless of whether the underlying data formats remain OGC-compliant. Here the policies-and-practices dimension, and with it the interoperability on which a standard’s public-good status depends, migrates to a trust-framework operator the SSOs do not control.

### *2.2.5 Comparative Insights and Lessons*

Taken together, these four cases illustrate that meso-institutional adaptation in voluntary standard-setting is neither uniform nor automatic. Each organization confronts different governance deficits (narrow participation, structural diffusion, participation bias, or jurisdictional fragmentation) and responds with different instruments (membership expansion, legal reconstitution, eligibility reform, or trust-framework development). What the cases share is a recognition that the internal governance architecture of the standard-setting institution is not a background condition but a material one, shaping whether the standards it produces can function as genuine public goods in an era of platformization and sovereign infrastructure. Across the four, the same dimensions recur: purpose, principles, people and processes, and policies and practices; and these dimensions help assessing whether an open standard meets the criteria of a digital public good (open licensing and access, interoperability, do-no-harm) or stays open only on paper.[21]

[20]Esri, “New Options for Decentralized Collaboration That Protects Data Sovereignty,” ArcNews, Winter 2025, https://www.esri.com/about/newsroom/arcnews/new-options-for-decentralized-collaboration-that-protects-data-sovereignty/. The article shows that ArcGIS architecture aligns with Gaia-X requirements, enabling decentralized collaboration that protects data sovereignty within the framework of European data space governance.

The comparative table below reveals that openness is not a binary property of a standard but a governance achievement produced across the 4Ps. Purpose determines whose public value standardization is intended to serve; Principles translate openness into non-negotiable commitments; People and Processes determine who can meaningfully shape those commitments; and Policies and Practices determine whether they survive implementation and institutional change. Together, these dimensions determine whether formal openness translates into the substantive characteristics of a digital public good—open licensing and access, interoperability, and do-no-harm—or remains openness largely on paper.

[21] Procedural openness is not sufficient: consensus can be 'procedurally precise but substantively hollow.' Marta Cantero Gamito, 'Harmonising Consensus,' 17 (2026) JIPITEC 75, 90.

Table 1. Governance challenges and reform responses across four organizations, read through the 4Ps

| Organization | Governance challenge | Purpose | Principles | People & Processes | Policies & Practices | DPG dimension most affected |
|---|---|---|---|---|---|---|
| **OGC** | Participation concentrated among well-resourced organizational actors; keeping standards aligned with contemporary implementation | Sustain geospatial standards as usable, public-interest infrastructure | Openness, inclusivity, interoperability | Individual Membership; sliding-scale fees; Agora; working groups, code sprints and testbeds broaden and lower barriers to participation | API modernization and Building Blocks; governance incorporated explicitly into institutional agenda | **Access & interoperability** |
| **W3C** | Diffuse authority and accountability under the former multi-host institutional structure | Preserve the Web as an open, globally governed infrastructure | Openness, accountability, transparency, public-interest governance | Member-elected Board majority and clearer strategic and fiduciary authority | Public-interest nonprofit structure; published bylaws; independent audits and financial disclosure | **Open governance & access** |
| **IETF** | Formal openness coexisting with persistent geographic and organizational participation inequalities | Maintain Internet protocols as globally legitimate shared infrastructure | Open participation, rough consensus, technical merit, global reach | Remote participation recognized in NomCom eligibility; proposed mentorship, multilingual onboarding, university engagement and community consultation | Formal rule reform combined with informal practices aimed at making participation practically — not merely formally — accessible | **Inclusive access & do-no-harm** |
| **Gaia-X** | Governance increasingly shifts from individual standards to trust frameworks, compliance mechanisms and infrastructure | Enable trusted, sovereign and interoperable data exchange across federated ecosystems | Trust, sovereignty, interoperability, discoverability, traceability | Governance extends beyond traditional SSO consensus processes to actors administering trust and compliance frameworks | Trust Framework, Data Exchange Services, trust labels and automated compliance become operative rules for data exchange | **Interoperability & governability** |

A further lesson cuts across all four cases. Work on public-interest participation in standard-setting dating back to the early 2000s, much of it developed by civil-society organizations with philanthropic support (see, e.g., Morris and Davidson 2003), identified the barriers that keep public-interest actors out of SSO processes even when the doors are formally open: technical expertise, resources and the need to prioritize where to engage, travel costs, organizational capacity, the difficulty of understanding complex procedures, and the demands of sustained engagement. Two decades later, the same barriers persist. Expanding participation therefore requires expanding the capacity to participate. The challenge is not only institutional openness but institutional intelligibility: governance bodies must invest in lowering the cognitive, financial, and organizational barriers to meaningful participation, through onboarding and mentorship, plain-language documentation of processes, fee structures calibrated to capacity, and collaboration tooling that makes decision pathways discoverable. Several of the reforms examined above are best understood in these terms. The documentary cases and the practice-derived observations disclosed in Section 1.4 therefore support treating intelligibility as a governance risk: formal openness can reproduce participation bias when the cognitive, financial, and organizational costs of participation remain high.

## 3. Platformization and the Sovereign Stack Challenge

### 3.1 Platform Enclosure of Geospatial Coordination

Traditional standardization conceives of geospatial standards primarily as documents: human-readable specifications for coordinate reference systems, feature models, encodings, and web services, adopted by SSOs and implemented by vendors (UN-GGIM 2022; OGC 2024). Governance in this view concerns who participates in drafting these documents, how consensus is defined, and what licensing conditions apply (Baron et al. 2019).

In contemporary practice, geospatial ecosystems have become increasingly API-centric. Cloud-based mapping, navigation, and earth-observation platforms expose data and functionality primarily through proprietary APIs and software development kits tightly integrated with identity, billing, logging, and machine-learning services (Kaleagasi 2022). Work on platform governance has shown that APIs encapsulate both technical and governance decisions: they define what can be done, who can do it, under what conditions, and how changes are communicated (Helmond 2015). In geospatial standards, this means that platform APIs can become de facto coordination rules for developers and institutions through their resource structures, default projections, event semantics, client libraries, and rate-limit policies.

Two features of this shift are worth naming, because together they explain why it is not self-correcting. The first is directional. Work on platform lifecycles describes a characteristic sequence of degradation, widely termed “enshittification” and taken up in the scholarly literature as platform decay: a platform is generous to users while it is accumulating them, then reallocates value from users to its business customers, and finally from those business customers to itself (Doctorow 2023; Ardoline and Lenzo 2025). The sequence matters here because it describes an equilibrium rather than misconduct: once a platform's interfaces have become the default against which others build, the cost to the platform of degrading them falls. The second feature is architectural. Research on smartphone infrastructures has shown that openness secured at one layer of a stack can be comprehensively neutralized by control exercised at another: the internet's founding commitments to openness and interoperability remain formally intact at the network layer, while what users and developers can actually do is governed at the device and operating-system layer, where those commitments do not reach (Veale 2026). Open geospatial standards face the same displacement one layer up, and for the same reason: the layer that determines practice is not the layer at which openness was secured.

Three cases illustrate how this platformization operates.

*Niantic's proprietary location stack.* Niantic's augmented-reality ecosystem operationalizes proprietary location anchors and discovery mechanisms that shape what counts as a usable location. The company's decision to sell its games business to Scopely in 2025 was analytically revealing: proprietary location primitives (point-of-interest layers, geofencing logic, community contribution tooling) were bundled with the product and transferred as an asset, along with decision rights over interfaces and data models. Subsequently, Niantic's geospatial business was spun off as Niantic Spatial Inc., and the migration from Lightship.dev to Scaniverse in early 2026 entailed the transfer of workspaces, POIs, scans, VPS maps, and API keys, with Lightship.dev decommissioned on February 27, 2026.[22] This lifecycle (deprecation, migration, platform reconstitution) shows how a private stack can unilaterally redefine the 'rules of place' through platform-managed decisions, outside any multi-stakeholder accountability process.

*Google Maps Platform.* Google Maps Platform shapes de facto norms for points of interest, geocoding, routing, and categorization through proprietary data models and APIs. The governance consequences became more visible in 2025, when Google restructured its pricing: the free monthly credit of $200 was terminated on March 1, 2025; Professional and Professional Advanced tiers were introduced on October 1, 2025.[23] The EEA-specific Terms of Service, effective July 8, 2025, required developers to use the new Places UI Kit when

---

[22] Niantic's platform governance is visible in its documented migration from legacy Lightship.dev to Scaniverse, including the transfer of sites, scans, and VPS maps and the deprecation timeline—an explicit example of how a private stack can change the operational 'rules of place' through platform-managed lifecycle decisions. Self-serve migration opened on February 20, 2026; Lightship.dev was decommissioned on February 27, 2026, with content migration required before February 27, 2027. See: Niantic Spatial Platform, "Migration Guide: Upgrade from Legacy Lightship.dev to Scaniverse," Niantic Spatial Platform documentation, 2026, https://www.nianticspatial.com/docs/scaniverse/migration_guide/. The geospatial business was spun off as Niantic Spatial Inc., and Scaniverse was launched on April 7, 2026, alongside VPS 2.0. See: Niantic Spatial, "Mapping the World for Machines with Scaniverse," Niantic Spatial Blog, 2026, https://www.nianticspatial.com/blog/scaniverse.

[23] Starting October 1, 2025, Professional and Professional Advanced tiers were added to the Google Maps Platform Core Services, and Google Maps Platform began charging for their usage. Additionally, as of March 1, 2025, Google terminated the free monthly credit of $200 USD and instead introduced free billable events for individual APIs. See: Globema, "Leverage New Google Maps Platform Licensing Methods," January 13, 2025, https://google.globema.com/2025/01/13/new-google-maps-platform-licensing-methods/; Cloudfresh, "Google Maps Changes 2025: What You Need to Know," 2025, https://cloudfresh.com/en/blog/google-maps-platform-changes-2025/.

displaying Places-related content with maps.[24] These pricing and policy decisions shape the practical economics of geospatial application development and, by extension, which platform's coordination rules become the default.

*Cloud-native Earth observation platforms.* At the analytics end of the stack, platforms used for large-scale Earth observation and geospatial AI—including Google Earth Engine and Microsoft's Planetary Computer—standardize practice through integrated catalogs, computation primitives, and access rules. Decisions at these layers, from spatial tiling schemes, feature schemas, temporal aggregation, to representations of uncertainty, shape what is feasible and convenient in downstream applications, effectively setting de facto standards even where SSO documents offer alternative representations.

The governance consequence is consistent across cases: the operative coordination rules are increasingly determined by platform roadmaps and service terms rather than the transparency, balanced participation, and accountability mechanisms associated with voluntary SSOs. Openness is not, by itself, a guarantee of public-interest outcomes: procedural consensus can conceal normative fractures and dilute substantive safeguards to the lowest common denominator, which is precisely why governance design, not openness alone, is decisive.[25] Because geospatial services are bundled with identity, billing, logging, and AI tooling, switching between platforms involves multi-layer switching costs. Implementation-stage power thus enables platform operators to function as de facto standard-setters.

Platform influence should not, however, be equated automatically with displacement of an open standard. The relevant continuum runs from complementarity, where a proprietary service implements or extends an open specification, through implementation dependence and de facto supplementation, to functional substitution, where third parties coordinate primarily against the platform interface and switching away becomes materially costly despite the continued availability of an open alternative.

These displacements are not, however, purely external. They are facilitated where, at the moment of implementation, the open specification is not the more usable, current, or readily supported. Section 2.2.1 explains the mechanism in the OGC case, where a standards body's outputs no longer align with how developers actually build and the coordination function migrates to whatever actor does offer implementable interfaces. Platform enclosure and governance weakness can therefore become mutually reinforcing: where the standards process leaves an implementation gap, the interface that fills it may become the default against which others coordinate. This is why the reforms proposed in Section 4 are directed at lifecycle and implementation monitoring rather than at the platforms themselves. Those are the levers that determine whether such a space opens in the first place.

---

[24] See https://developers.google.com/maps/comms/eea/geocoding.

[25] Marta Cantero Gamito, 'Harmonising Consensus,' 17 (2026) JIPITEC 75, 90-93 (procedural consensus 'procedurally precise but substantively hollow'); cf. the 'non-participatory turn' in CEN-CENELEC's JTC21 working method (Oct. 2025) noted in the special issue's Editorial, 17 (2026) JIPITEC 1, 9.

## 3.2 Sovereign Cloud and Sovereign AI as Full-Stack Governance Projects

The emergence of sovereign-cloud and sovereign-AI initiatives intensifies these dynamics rather than resolving them. Digital sovereignty is the frame within which these initiatives are advanced; sovereign cloud and sovereign AI are the forms in which it becomes infrastructure, and it is in that infrastructural form that it bears on standards governance. The distinction matters because the frame is contested in a way the infrastructure is not: sovereignty as a political objective and sovereignty as an operational capability come apart, and governments pursuing the first frequently remain dependent on the providers that supply the second. Sovereign-cloud strategies seek to reconcile cloud computing with legal and political requirements for data protection, cybersecurity, and strategic autonomy (OECD 2024, Dilanyan 2026). Meaningful sovereignty requires control not only over data location but also over who operates data centers, who has access to operational telemetry, and how updates, patches, and incident responses are governed. Control over APIs and management plans becomes a key aspect of sovereignty (Blancato 2023, Cruzes 2026, Delima 2026).

In Europe, these dynamics are particularly visible. Gaia-X has evolved into an operational framework with trust labels indicating levels of compliance, security, and data governance, with higher trust levels positioned as compatible with forthcoming EUCS "High+" requirements (Gaia-X 2025). The global sovereign-cloud market is projected to grow from approximately $195 billion in 2026 to over $1.1 trillion by 2034 (Fortune Business Insights 2026). Gartner has named geopatriation, defined as the strategic migration of workloads from global public clouds to local or sovereign environments, as a top strategic technology trend for 2026 (Gartner 2025). These developments are relevant here not as evidence that sovereign infrastructure is necessarily more closed, but because they move questions of interoperability, control, and exit into the architecture of the infrastructure itself.[26]

For geospatial standardization, sovereign-cloud and sovereign-AI architectures embed legal and policy requirements into infrastructure stacks, data pipelines, and AI toolchains, transforming cloud and AI architectures into explicit governance projects. When geospatial data and services are deployed within sovereign-stack environments, the governance rules that apply are determined not only by the SSO standards they nominally implement but also, and often primarily, by the infrastructure operator's compliance frameworks, API management policies, and service terms. If the sovereign-stack operator is a hyperscaler offering a "sovereign" partition of its global cloud, the governance rules may be formally responsive to national law while remaining operationally determined by the platform's architecture and business model. Work on anticipatory data governance draws the distinction that matters here, between sovereignty as a political objective and sovereignty as an operational capability: governments pursuing autonomy frequently remain dependent on foreign cloud providers, platform infrastructure, and technical services, and sovereignty can become a reactive posture that prioritizes restriction before addressing the underlying questions of governance, capability, incentives, and public value (Zable and Verhulst 2026). That architecture is consequential in law and not only in practice. Comparative work on the informational infrastructures built during the Covid-19 pandemic has shown that the choice between centralized and decentralized system architectures determined which legal instruments could reach those systems at all and what courts could effectively review; architecture set the boundaries of legal control rather than merely operating within them (Veale 2024).

---

[26] Gartner Insights Abstract, 18 October 2025. https://www.gartner.com/en/documents/7079998.

The governance question is therefore less whether platformization simply advances or retreats than where coordination authority comes to rest and on what terms. Contemporary software development gives the implementation layer substantial practical weight, but the degree of displacement remains an empirical question in any particular market. The normative question is whether authority at that layer remains contestable: whether rules are disclosed, whether those bound by them can challenge them, whether interoperability is maintained, and whether exit is possible at tolerable cost. Framed this way, digital-sovereignty initiatives may operate either as a counterweight to concentrated platform power or as a relocation of it. Jurisdictional compliance is therefore not, by itself, evidence of governable openness.

Sovereign-cloud and sovereign-AI initiatives should consequently be evaluated against contestability and portability rather than sovereignty labels alone. They may strengthen openness where federated governance, transparent rules, interoperable interfaces, and credible exit reduce dependency. They may intensify platformization where national legal requirements and certification regimes are superimposed on provider-specific APIs and infrastructure without restoring meaningful choice. The issue is thus not the presence of an additional governance layer as such, but whether that layer makes coordination authority more transparent, contestable, and portable.

## 3.3 Risks for FAIR Data, Digital Twins, and the Industrial Metaverse

The platformization and sovereign-stack dynamics described above do not operate in isolation. They interact with—and threaten to undermine—a set of foundational commitments that the geospatial community has developed over the past decade to ensure that geospatial data serves broad public and innovation purposes. Three domains show what is at stake, and each shows something different. In FAIR data practice, the risk is that formal compliance and practical openness come apart. In digital twins, it is that interoperability foundations fragment along jurisdictional and platform lines even where the technical standards remain compatible. In the industrial metaverse, where the relevant standards are not yet settled, it is that the operative spatial representations are fixed by implementation before open standardization reaches them.

*The FAIR principles and geospatial data access for innovation.* The FAIR principles—Findability, Accessibility, Interoperability, and Reusability—have become a foundational framework for data governance in scientific, governmental, and increasingly commercial contexts. In the geospatial domain, OGC has long supported FAIR-compliant data practices and the NSDI Strategic Plan 2025–2035 explicitly commits to ensuring that spatial data follows FAIR principles while embracing innovations such as artificial intelligence and digital twins (FGDC 2025). OGC's RAINBOW initiative provides a web-accessible source of information about concepts and vocabularies that apply FAIR principles to key concepts underpinning interoperability, while the proposed Sample Markup Language for AI/ML would enable users to document, store, and share geospatial sample data following FAIR data management principles (OGC 2025).

FAIR enters this analysis as an object of study rather than as a premise. It is the vocabulary in which the geospatial community has stated its own commitments, through IGIF, the UN-GGIM/OGC Standards Guide and the NSDI Strategic Plan among others, and the question we pursue is whether those commitments survive implementation, not whether they are the right commitments to have made.

Platformization can weaken the practical realization of each FAIR dimension even when formal alignment with FAIR remains possible. Findability is compromised when proprietary catalogs and discovery mechanisms replace open metadata registries: if geospatial datasets are discoverable only through platform-specific search interfaces, they become findable only within the platform ecosystem, not across the broader data landscape. Practical accessibility is undermined when APIs impose rate limits, tiered pricing, or authentication requirements or other conditions that make access dependent on a particular commercial relationship or platform environment. Such conditions do not necessarily defeat formal FAIR accessibility, which can accommodate authenticated or conditional access, but they can materially affect who can access data in practice and on what terms: Google Maps Platform's 2025 pricing restructuring illustrates how changes in access economics can reshape who can practically reach geospatial data and services. Interoperability—the dimension most directly connected to standards governance—is eroded when platform-specific data models, projections, and feature schemas become the operative representations, even when nominally compatible with open formats at the margins. Reusability is diminished when licensing terms, usage restrictions, and proprietary provenance mechanisms constrain how geospatial data can be combined, transformed, and repurposed downstream.

The aggregate effect is best understood as a form of practical FAIR erosion operating through architectural displacement rather than explicit rejection of the principles: the infrastructure within which geospatial data is produced, stored, and exchanged increasingly conditions how FAIR characteristics can be realized in practice. This need not entail formal non-compliance with FAIR; a platform may satisfy formal FAIR criteria while effective openness, portability, and contestability come apart. For data-driven innovation, including the training and validation of geospatial AI models, the construction of integrated spatial analytics, and the development of cross-domain applications, this erosion is consequential. Innovation ecosystems depend on the ability to combine heterogeneous data sources in unanticipated ways; FAIR principles provide a foundational governance framework that makes such combinatorial innovation possible. When that framework is undermined by platform-specific constraints, the innovation space contracts.

*Digital twins and the interoperability imperative.* Digital twins represent one of the most significant contemporary applications of geospatial data and standards. Urban digital twins—virtual replicas of cities integrating infrastructure, environmental, and social data for planning, management, and decision-making—depend fundamentally on the interoperability of heterogeneous data sources at multiple spatial and temporal scales (Gonzalez Otero and Verhulst 2025). The OGC Urban Digital Twins Discussion Paper (2024) identifies this dependency explicitly, positioning OGC standards (CityGML 3.0, OGC SensorThings API, OGC API – 3D GeoVolumes, OGC API – Features, WFS, WMS) as enabling components of scalable and interoperable digital twin ecosystems.[27] The Urban Digital Twins Interoperability Pilot aims to enhance geospatial data interoperability and increase use of open standards within urban digital twin implementations.

The analytical challenge is that digital twins are, by their nature, integrative systems. They must ingest and harmonize data from built-environment sensors (IoT), satellite and aerial imagery, cadastral and land-use databases, transportation networks, utility systems, and environmental monitoring stations. Each of these data domains has its own standards, formats, and governance arrangements. CityGML 3.0, for instance, standardizes the underlying information model for 3D city models and offers significantly improved integration with Building Information Modelling (BIM), support for indoor spaces at different Levels of Detail, and the capacity to represent dynamic data from sensors and simulations. But the practical value of these standards depends on the governance infrastructure that sustains them: if the effective data-exchange rules in a given digital twin deployment are determined by a proprietary platform's APIs and data models rather than by OGC specifications, then the interoperability promise of the standard is formally maintained but practically hollow.

Sovereign-cloud dynamics compound this challenge. When digital twin deployments operate within sovereign-stack environments, the data-governance rules that apply—access controls, data-residency requirements, compliance certifications—are set by the sovereign-stack operator, not by the geospatial SSO whose standards the twin nominally implements. If different sovereign-cloud environments impose different governance rules on the same underlying data and standards, the result is a fragmentation of digital twin interoperability along jurisdictional and platform lines, even where the underlying technical standards remain compatible. Current research confirms that data interoperability, system integration, and scalability remain the primary challenges in urban digital twin development (Schrotter and Hürzeler 2020; Shahat, Hyun, and Yeom 2021). Platformization and sovereign-stack fragmentation risk making these challenges structural rather than merely technical.

*The industrial metaverse and spatial computing.* The implications extend beyond urban digital twins to the broader domain of the industrial metaverse and spatial computing—areas where the geospatial standards stack serves as a foundational layer for emerging technology ecosystems. The Metaverse Standards Forum, established in 2022 with over 1,200 member organizations, operates ten Domain Groups and three Exploratory Groups covering 3D interoperability, digital twins, and geospatial systems.[28] OGC is a founding member, and the OGC Geo for Metaverse Domain Working Group has been established to coordinate the integration of geospatial standards within the metaverse, including compilation of existing standards and definition of requirements for fostering an open and interoperable metaverse. The OGC GeoPose standard provides geoanchored location and orientation information in a developer-friendly format. The Alliance for OpenUSD is working to make OpenUSD perform well for real-time, large-scale virtual worlds and industrial digital twins, including building support for geospatial coordinates and real-time streaming of IoT data.

[27]Open Geospatial Consortium, "Urban Digital Twins: Integrating Infrastructure, Natural Environment, and People," Discussion Paper, OGC 24-025, 2024, https://docs.ogc.org/dp/24-025.html. See also Chester, Simon , "Urban Digital Twins: Planning the Cities of Tomorrow," OGC Blog. March 2021. https://www.ogc.org/blog-article/urban-digital-twins-planning-the-cities-of-tomorrow/. The Urban Digital Twins Interoperability Pilot (UDTIP) aims to enhance geospatial data interoperability and increase use of geospatial standards within Urban Digital Twins, using OGC SensorThings API, OGC API – 3D GeoVolumes, OGC CityGML, OGC API – Features, WFS, and WMS.

The industrial metaverse presents a particularly acute case of the governance-infrastructure gap. Industrial metaverse applications such as factory digital twins, infrastructure lifecycle management, energy-system modeling, and logistics optimization, require the integration of geospatial data, BIM data, IoT sensor streams, and simulation outputs within persistent, shared, and interoperable virtual environments (IEC 2025). The GeoBuiz Summit 2026 identified "Spatial Computing and Digital Twin Enterprise" as the central driver of digital transformation across industries.[29] But achieving this integration at scale requires that the underlying standards[30] are governed through processes that ensure long-term interoperability, vendor neutrality, and the capacity for combinatorial innovation.

Platformization threatens this integration in a specific way. If the operative standards for 3D geospatial representation, scene description, and spatial anchoring are determined by platform-specific implementations rather than by open SSO processes, then the industrial metaverse fragments along platform lines: industrial actors locked into one platform's spatial representations cannot seamlessly interoperate with actors on another. The multi-layer switching costs (data migration, re-implementation of application logic, staff retraining) are multiplied when the standards at stake are not merely data-exchange formats but entire spatial computing environments. The governance stakes are correspondingly higher: the standards that govern how physical space is represented, indexed, and operated upon in the industrial metaverse will shape industrial organization, supply-chain architecture, and competitive dynamics for decades.

---

[28]OGC joined the Metaverse Standards Forum as a founding member in 2022. See: Spatial Source. "OGC Joins New Metaverse Standards Forum." July 2022. https://www.spatialsource.com.au/ogc-joins-new-metaverse-standards-forum/. The Geo for Metaverse Domain Working Group has been established to coordinate the integration of geospatial standards within the metaverse, including compiling existing standards from OGC and beyond, and defining requirements for fostering an open and interoperable metaverse. See: OGC, "Geo for Metaverse Domain Working Group," https://www.ogc.org/about-ogc/committees/dwg/geo-for-metaverse-domain-working-group/. See also: Chester, Simon. "The Metaverse Is Geospatial." OGC Blog. May 2022. https://www.ogc.org/blog-article/the-metaverse-is-geospatial/.

[29]GeoBuiz Summit 2026 (Denver, Colorado, January 12–14, 2026) had as its theme "Spatial Computing and Digital Twin Enterprise," spotlighting how geospatial technology is evolving to become a central driver of digital transformation across industries. See also: IEC (International Electrotechnical Commission), Industrial Metaverse: Standards and Technology Report (Geneva: IEC, 2025), https://www.iec.ch/system/files/2025-10/iec_sttr_industrial_metaverse_en_lr_1.pdf.

[30] These underlying standards are for 3D geospatial representation (CityGML, 3D Tiles, I3S), for sensor data (SensorThings API), for pose and orientation (GeoPose), and for scene description (OpenUSD, glTF).

**Table 2. How platformization and digital sovereign-stack dynamics bear on each FAIR dimension**

| Example | Findable | Accessible | Interoperable | Reusable | Principal FAIR risk |
|---|---|---|---|---|---|
| **Niantic / proprietary location stacks** | Location objects, POIs and spatial anchors are discoverable primarily within proprietary platform environments rather than through shared registries. | Access depends on platform APIs, accounts and continued platform support. | Proprietary location primitives, geofencing logic and interfaces can become the operative spatial model, creating dependencies beyond open standards. | Migration, deprecation and transfer of platform assets can change the conditions under which spatial resources remain usable downstream. | **Platform dependency across the lifecycle** |
| **Google Maps Platform** | Places and geospatial resources are increasingly discovered through Google-controlled catalogs and interfaces. | Pricing tiers, authentication, quotas and service terms determine practical access, as the 2025 pricing restructuring illustrates. | Proprietary Places models, APIs and required interface components can become de facto coordination rules even where open standards remain available. | Terms of service and restrictions on how platform-derived content can be combined or displayed constrain downstream reuse. | **Accessibility and interoperability governed by commercial terms** |
| **Cloud-native Earth-observation platforms** | Integrated platform catalogs make data highly findable within the platform, but can make discovery increasingly platform-specific. | Data and computation are accessed through platform-managed accounts, APIs and compute environments rather than independently of them. | Tiling schemes, feature schemas, temporal aggregation and uncertainty representations become de facto standards for downstream workflows. | Reuse becomes coupled to platform computation primitives, provenance systems and workflows, increasing switching costs. | **Architectural FAIR erosion by platform defaults** |
| **Gaia-X / sovereign data-space infrastructures** | Federated catalogs and trust frameworks can potentially strengthen discovery across data spaces. | Access becomes conditional on trust labels, compliance rules and sovereign-infrastructure requirements. | Gaia-X promotes interoperability, but effective interoperability increasingly depends on compatibility with its trust framework and compliance layer as well as underlying OGC/ISO standards. | Provenance and usage conditions can become more structured, but reuse may become conditional on compliance and governance rules at the data-space layer. | **FAIR becomes mediated by an additional governance layer** |
| **Sovereign cloud / sovereign-AI stacks more broadly** | Discovery may fragment across jurisdictional or provider-specific environments. | Access becomes governed not only by open standards but by sovereignty requirements, identity, certification and API policies. | Nominal standards compatibility may coexist with operational dependence on provider-specific infrastructure and APIs. | Reuse can be constrained by jurisdictional rules, platform architecture and service terms even when underlying data formats remain open. | **Formal FAIR compliance without practical portability** |

As the comparative table above indicates, for all domains, FAIR data practices, digital twins, and the industrial metaverse, the conclusion is the same. The risks of platformization and sovereign-stack fragmentation are not limited to the geospatial standards stack in isolation. They spread through every technology domain that depends on interoperable geospatial foundations. Open geospatial standards are not merely one set of technical agreements among many; they are *enabling infrastructure* for a wide range of frontier-technology ecosystems. When the governance of those standards is compromised, whether through platform capture, sovereign-stack fragmentation, or meso-institutional weakness, the effects cascade across the entire innovation landscape. Across all three domains, the feedback mechanism identified in Section 1.3 can arise where an open specification is not the more implementable option at the moment of choice: coordination authority may migrate toward the interface or infrastructure that is, and accumulated implementation dependencies and switching costs can make that migration difficult to reverse. The downstream significance of this mechanism depends on where implementation sits on the continuum identified in Section 3.1: complementarity need not threaten open coordination, whereas implementation dependence and functional substitution increasingly relocate effective authority to the implementation layer.

## 4. The 4Ps Framework for Governable Geospatial Standards

The preceding analysis isolates the governance choices that materially condition the capacity of an open geospatial standard to function as a digital public good. We consolidate them into a framework of four interdependent dimensions: Purpose, Principles, People and Processes, and Policies and Practices (the 4Ps), adapted from data-governance scholarship and calibrated here to voluntary standards-setting organizations.[31] The framework is both diagnostic, locating where a given SSO is strong or exposed, and prescriptive, indicating the reforms a body must undertake to remain a credible steward of open standards. Crucially, the dimensions describe a continuing practice, not a one-time audit: openness is something an institution must keep earning. It is a diagnostic framework rather than an empirical claim that particular failure modes occur with a given frequency across SSOs; its categories synthesize the documentary cases, established governance literature, and the practice-derived questions disclosed in Section 1.4.

[31] The 4Ps build on Stefaan Verhulst, Data Governance Toolkit: Navigating Data in the Digital Age (Broadband Commission for Sustainable Development, 2025) (mapping Purpose/Principles/People and Processes/Policies and Practices to WHY/HOW/WHO/WHAT); and Gonzalez Otero and Verhulst, Cities and Digital Twins (2025).

Table 3. The 4Ps as a diagnostic framework

| Dimension | What it asks | What it demands | Diagnostic question |
| --- | --- | --- | --- |
| **Purpose (WHY)** | What public value the standard and its steward exist to create, and for whom | A stated public-interest mission; identified communities that depend on the standard as infrastructure; a roadmap that serves them rather than the largest members | **Whose public value does this standard serve, and can the SSO show it?** |
| **Principles (HOW)** | Which commitments are non-negotiable | Open licensing and free public accessibility of standards that function as infrastructure; interoperability across the whole stack; do-no-harm | **Are the standard's public-good criteria guaranteed, not just aspired to?** |
| **People and Processes (WHO)** | Who participates and how decisions are made and contested | Affordable, inclusive membership and access; balanced, disclosed working-group leadership; appeal and revision pathways; insulation from platform and incumbent capture; an accountable stewardship function; documented decision responsibilities and a way to declare interests | **Can those affected by the standard meaningfully shape it?** |
| **Policies and Practices (WHAT)** | What the organization actually does over time | Lifecycle maintenance; monitoring of implementation, including platform APIs that may displace the standard; periodic review against purpose; transparent reporting; clear triggers for retiring obsolete standards | **Is openness sustained in practice, or only on paper?** |

Purpose failures usually take the form of drift. As a standards body's activities expand beyond their original remit, from standard-setting into research, convening, and increasingly policy engagement, its mission is rarely revisited and no routine exists for reviewing or updating it. The result is fragmented initiatives, misaligned incentives, and a culture that struggles to decline work, diluting focus and overloading limited governance capacity. The proposed remedies are structural but light-touch: codify the body's main program areas in its constitutive documents, institutionalize a periodic strategic-governance review, and apply an explicit mission-fit test before new work is taken on.

Principles risks arise where commitments such as openness, consensus, interoperability, and trust remain aspirational rather than operational. Standards may be reusable and interoperable yet difficult to find; transparency may depend on tacit knowledge rather than documented process; and public-interest or ethical implications may receive no structured consideration even where technically sound specifications generate societal risks. The response is to consolidate principles into a published set, embed them in charters, templates, and review rubrics, and add a lightweight Public Interest Test across the standards lifecycle. This is not the first proposal to introduce public-interest questions into technical standardization: Morris and Davidson (2003) proposed a public-policy impact assessment for Internet standards. The contribution here is more specific to the present argument. It embeds the inquiry at successive lifecycle decision gates and adds contemporary questions of platform dependency, contestability, infrastructure concentration, and digital-public-good stewardship. Existing standards processes commonly foreground market demand, technical consensus, and implementation experience; the proposed test adds an explicit public-interest judgment and requires that it be revisited as implementations, dependencies, and usage contexts change. The test consists of a short set of questions grouped along four dimensions:

**Table 4. The Public Interest Test**

| Public Interest Test | Questions to ask at each decision gate of the lifecycle |
|---|---|
| ***Distribution of benefits and risks*** | – Who benefits?<br>– Who bears the risks?<br>– Which stakeholders were absent from the process? |
| ***Societal impact and do-no-harm*** | – Does the standard raise ethical or societal concerns, such as privacy, equity, data sensitivity, or environmental impact?<br>– Have these concerns been documented and, where needed, mitigated? |
| ***Market structure and dependency*** | – Does the standard increase interoperability?<br>– Does it increase contestability?<br>– Does it reduce, or deepen, dependency on particular infrastructures?<br>– Does it create disproportionate advantages for particular platforms? |
| ***Future orientation and public value*** | – Does it preserve space for future innovation?<br>– Does it support Digital Public Goods principles (Digital Public Goods Alliance 2025)? |

What makes the test consequential is where it applies. A standard's lifecycle is marked by decision gates, and each gate changes the meaning of the questions. At chartering, the test screens who is at, and who is absent from, the table before positions harden. This screening is a documentation duty on the proposing group — a short, recorded answer to who is affected and not represented — rather than an obligation to convene constituencies beyond the membership; the published record then does its work through the public-comment, liaison, and review channels that consensus processes already provide. When early-stage specifications graduate from incubation into formal standardization, it checks whether experimentation has already created de facto dependencies. When research and pilot outputs transition into the standards track, the point at which sponsor-driven agendas most need a public-interest check, it documents whose interests the transition serves. At adoption, it verifies that concerns recorded at earlier gates were resolved rather than merely noted. After adoption, through light-touch post-adoption monitoring, it re-examines standards whose usage, dependencies, or societal context has shifted, rather than presuming them benign. And at retirement, the decision to sunset a legacy standard looks different once "who bears the risks?" includes the public agencies that depend on it as critical infrastructure for decades. At each gate, short, published answers form part of the decision record rather than a new approval layer, keeping the test lightweight enough for volunteer-driven organizations.

The proposal has antecedents that should be named. Structured public-interest assessment for technical standards was put to the IETF two decades ago as a set of public-policy questions to accompany standards development (Morris and Davidson 2003), and public-interest review embedded in a standard-setting lifecycle is established practice in other domains, including international assurance standard-setting conducted under public-interest oversight. What we add is narrower: the questions are calibrated to voluntary technical SSOs, they are asked at every decision gate rather than once at chartering, and they incorporate concerns the earlier proposals could not anticipate, namely platform dependency, contestability, infrastructure concentration, and digital public goods stewardship. This lifecycle embedding is what distinguishes the proposal from the one-off ethics or impact assessments occasionally discussed in standardization (see, e.g., Casiraghi and van Dijk 2026): a single review at chartering cannot see the harms, dependencies, and exclusions that only materialize in implementation. The test does not resolve what the public interest requires, and it is designed not to. The distinction drawn in the early critical literature on computing between deciding and choosing is suitable here: instrumental reason can make decisions, but deciding and choosing are not the same act (Weizenbaum 1976, 259). Deciding is the kind of activity that can be delegated to a procedure; choosing is the product of judgment and cannot be. A public-interest test rendered as a scoring rubric would convert the second into the first and, in doing so, discard precisely what needed to be judged. What the test does instead is ensure that a judgment is made rather than defaulted into: that it is made by identified people, at a named point in the lifecycle, on a published record that those affected can contest. Its output is an attributable choice, not a score, and it turns the throughput transparency discussed in Section 2 into routine practice. The societal-impact questions anchor the test in the do-no-harm criterion that, alongside open licensing and interoperability, defines a digital public good: as the history of web cookies illustrates, technically sound specifications can carry societal consequences that no one was asked to consider. A negative or uncertain answer would not veto standardization: it would trigger a documented rationale and, where warranted, targeted consultation with the constituencies the test has identified as absent, thereby feeding back into input legitimacy. The questions on contestability and dependency, in turn, link the test directly to the sovereign-stack dynamics analyzed in Section 3: they operationalize, at the level of individual standards decisions, the concern that openness can be hollowed out by the infrastructures on which implementation depends.

People-and-processes risks include concentration, opacity, and low institutional intelligibility. Procedural authority and institutional memory can become concentrated in a small number of individuals; participation costs can privilege well-resourced incumbents; and poorly documented responsibilities can make accountability difficult to locate. Sponsor-funded or experimental work may also sit at the edge of formal oversight. Clear allocations of roles and responsibilities, harmonized voting and approval rules, published decision templates, mechanisms for declaring interests, and deliberate onboarding and mentoring can convert informal practice into more accountable process. These investments are what make input legitimacy practical rather than merely formal.

Finally, openness is sustained or lost in the policies and practices that surround a standard over its life. Risks arise when policy landscapes become difficult to navigate, onboarding remains costly, dispute resolution depends primarily on informal relationships, or lifecycle decisions lack clear criteria for maintaining and retiring standards on which different constituencies depend. Potential responses include a public lifecycle dashboard, an incubation track for early-stage specifications, light-touch post-adoption monitoring, and formal dispute mechanisms that complement rather than replace consensual culture. This dimension carries the weight of the feedback mechanism proposed in Section 1: lifecycle currency and implementation monitoring can help keep an open specification a credible implementation option; where they lapse, coordination authority may migrate elsewhere, although that causal sequence remains an empirical question rather than an assumption of the framework.

Applied to the four cases examined above, the 4Ps convert diffuse observations into a structured account: the individual-membership and nonprofit-reconstitution reforms act mainly on People and Processes and on Principles; the participation-diversity reforms on People and Processes; and trust-framework development on Policies and Practices, at the contested boundary where stewardship migrates beyond the standards body itself. Read together, the dimensions show why technical openness alone cannot secure a digital public good, and what governance, sustained over time, would be required to do so. The 4Ps also give the three legitimacy dimensions introduced in Section 2 institutional traction: People and Processes largely determine input legitimacy; Principles, together with the procedural elements of People and Processes, sustain throughput legitimacy; and Purpose and Policies and Practices are where output legitimacy is earned or forfeited. In this sense, the framework operationalizes for standards bodies the legitimacy criteria on which their continued authority depends.

The four dimensions describe instruments internal to a standards body and that is what makes them actionable: they are levers a body can pull without waiting for anyone else. Where coordination authority has already migrated to platform interfaces, service terms, and trust frameworks, a standards body can monitor that migration, document it, and decline to ratify it. Governing it directly calls on instruments held elsewhere: the legal conditions under which a compatible implementation may be built, the competition-law treatment of an incumbent whose interface has become the default, and the procurement and recognition decisions of public authorities. Distinguishing the two is part of what the framework offers. It sets out what a standards body can secure on its own, and identifies where the remaining conditions sit and who holds them, which is where engagement beyond the standards body would begin. The distinction is the one drawn at the outset of this paper: a well-formed internal process is a condition of governable openness, in the way that an open license is a condition of a public good, indispensable, and doing its work alongside the rest.

## 5. Conclusion: Toward Governable Openness

Our main argument can be refined into a single proposition: the capacity of open geospatial standards to function as digital public goods, and once embedded in public systems as components of digital public infrastructure, depends not on the technical openness of specifications alone but also on the governance capacity of the institutions that produce and steward them and by the infrastructures through which they are implemented. The policy conclusion is accordingly not that openness should be equated with governance, but that any adequate account of openness must include governability.

The first claim, concerning the internal governance architecture of SSOs understood as meso-institutions, reveals that these organizations perform a critical translation function: they convert broad normative commitments (openness, interoperability, public interest) into the specific procedural, technical, and organizational arrangements that make those commitments operational. The quality of this translation depends on governance design choices that are often treated as peripheral: who can participate, how decisions are made, how conflicts of interest are managed, how the lifecycle of standards is stewarded, and how accountability is ensured. The cases examined illustrate that meso-institutional adaptation takes diverse forms but responds to a common pressure: the need to demonstrate that SSO governance is not only procedurally legitimate but operationally capable of sustaining broad participation, timely decision-making, and credible lifecycle stewardship.

The second claim, concerning the platformization intensified by digital sovereignty dynamics, shows that the effective governance of geospatial standards increasingly occurs outside SSO processes, in platform roadmaps, API management decisions, pricing structures, and sovereign-stack compliance frameworks. The risks are not confined to the geospatial domain: they propagate through every technology ecosystem that depends on interoperable and open geospatial foundations, from FAIR-compliant data practices and digital twins to the industrial metaverse and spatial computing. The feedback loop between platformization and SSO governance weakness creates the conditions for a self-reinforcing dynamic in which reduced engagement in open standardization further weakens collective governance capacity, progressively undermining the essential infrastructure character of open geospatial standards.

Our conclusions are informed not only by the framework developed above but also by advisory work with an open geospatial standards body on its internal governance architecture, which allowed the design choices described here to be examined in operation: how membership structures shape participation patterns, how decision-making processes allocate influence, how lifecycle stewardship affects the timeliness and relevance of outputs, and how accountability mechanisms, or their absence, condition an organization's legitimacy and its attractiveness to potential members. Two observations from that work have been carried into the framework rather than left beside it. The capacity for strategic self-reflection, and the ability to articulate and operationalize a public-interest mission, belong to Purpose, and are the reason that dimension calls for a periodic governance review rather than a one-time mission statement. The effectiveness with which an organization communicates the value proposition of open standardization to constituencies that might otherwise default to platform-provided alternatives belongs to Policies and Practices, and is a condition of the output legitimacy on which continued participation depends.

As anticipated earlier, the stakes of these governance choices and legitimacy demands are rising, not falling, in the AI era. Historically, standards mostly coordinated products; today they increasingly coordinate AI models, cloud infrastructure, digital identity, geospatial AI, digital twins, and public infrastructure itself. In doing so, they perform quasi-regulatory functions: they allocate rights, obligations, and market access. The clearest contemporary illustration is the role of harmonised standards under the EU AI Act, where technical specifications developed by standards bodies confer a presumption of conformity with binding legal requirements; formally voluntary instruments that acquire quasi-mandatory force, as the literature on transnational private regulation discussed in Section 2 anticipated, with well-documented consequences for transparency, fundamental-rights protection, and the political economy of who writes the rules (Volpato, Eliantonio, and Finck 2026; Cantero Gamito 2026; Eliantonio 2026; Oliva 2026). It is a regime whose own soft-law instruments are already proving contested in application (Veale and Quintais 2026). Where technical and soft-law instruments carry this much regulatory weight, the legitimacy of the bodies producing them ceases to be a supplementary consideration and becomes a condition of the scheme's validity. One asymmetry should be allowed here. Harmonized standards under the AI Act are produced by recognized European standardization bodies, not by the consortium organizations examined in this paper, and those bodies operate under participation and transparency obligations from which consortia are largely exempt. The asymmetry is the argument rather than an objection to it: the legitimacy demands now being made of recognized bodies are arriving at consortium bodies without the legal scaffolding that would make them enforceable there. Where law does not supply the constraint, internal governance must. Once standards begin to allocate rights, obligations, and market access, technical legitimacy is no longer sufficient: the input, throughput, and output legitimacy of the institutions that produce them becomes a condition of their continued authority.

An important policy implication is that open geospatial standards should be recognized and treated as candidate digital public goods and, once deployed in public systems, as digital public infrastructure, and the SSOs that produce them as the institutions that steward both. None of the three is reducible to the others, and none is merely a technical convenience. An open standard whose governance is opaque, whose participation is narrow, whose lifecycle is poorly managed, and whose implementation is effectively captured by platform operators is open only in a formal, documentary sense. Genuine openness requires governability: the capacity for the institutions that produce and maintain standards to sustain broad participation, transparent decision-making, responsive adaptation, and credible accountability across the full stack, from canonical standards and specifications to APIs, compliance frameworks, and implementation guidance.

Meeting the UN standards for geospatial information management requires not only technically sound standards but institutionally robust governance. The governance reforms examined in this paper represent steps in the right direction. But they will be insufficient unless accompanied by a broader reconceptualization of what it means for geospatial standards to be "open."

If the internal governance of geospatial SSOs can be strengthened along the lines suggested by the meso-institutional framework and future-proofed with the 4Ps framework set out above, these organizations can reinforce their value proposition as coordination points for an increasingly complex standards landscape and ensure that the standards they produce continue to serve the public interest in an era of platformization and sovereign infrastructure politics. The alternative, allowing the governance gap to widen, risks consigning open geospatial standards to formal existence without practical authority, and ceding the effective governance of location data to platform operators and sovereign-stack providers whose institutional logics may not be aligned with the public-interest commitments associated with digital public goods and digital public infrastructure. As standards increasingly shape public infrastructure, AI ecosystems, and sovereign digital capabilities, legitimacy becomes as important as technical excellence. Organizations that cannot demonstrate that their governance reflects the interests of the broader communities affected by their standards risk losing not only trust but, ultimately, their authority. Future standards-setting organizations must therefore evolve from communities of technical experts into institutions capable of sustaining public legitimacy.